\documentclass[12pt]{iopart}
\usepackage{iopams}  
\usepackage{setstack}
\usepackage{graphicx}
\usepackage[applemac]{inputenc}
\begin{document}
%------------------------------------------------------------------------------------------------------------------------------------------
\title{From mindless mathematics to thinking meat?}
%------------------------------------------------------------------------------------------------------------------------------------------
\author{Matt Visser}
%------------------------------------------------------------------------------------------------------------------------------------------
\address{School of Mathematics and Statistics\\
Victoria University of Wellington, PO Box 600, \\
Wellington 6140, New Zealand}
%------------------------------------------------------------------------------------------------------------------------------------------
\ead{matt.visser@sms.vuw.ac.nz}
\begin{abstract}
Deconstruction of the theme of the 2017 FQXi essay contest is already an interesting exercise in its own right:
Teleology is rarely useful in physics --- the only known mainstream physics example (black hole event horizons) has a very mixed score-card --- so the ``goals'' and ``aims and intentions'' alluded to in the theme of the 2017 FQXi essay contest are already somewhat pushing the limits. Furthermore, ``aims and intentions'' certainly carries the implication of consciousness, and opens up a whole can of worms related to the mind-body problem. 
As for ``mindless mathematical laws'', that allusion is certainly in tension  with at least some versions of the ``mathematical universe hypothesis''. 
Finally ``wandering towards a goal'' again carries the implication of consciousness, with all its attendant problems.

In this essay I will argue, simply because we do not yet have any really good mathematical or physical theory of consciousness, that the theme of this essay contest is premature,  and unlikely to lead to any resolution that would be widely accepted in the mathematics or physics communities.

\bigskip
\noindent
{\it Keywords}:  Mathematical physics; thinking meat; mind-body problem; consciousness; quantum; teleology; ontology; epistemology.

\vskip 10 pt
\noindent 
1 March 2017; 27 July 2917; \LaTeX-ed \today.

\vskip 10 pt
\noindent 
Essay written for the FQXi 2017 essay contest: ``Wandering towards a goal: \\
How can mindless mathematical laws give rise to aims and intention?''

\end{abstract}

%\pacs{89.70.Cf;  89.70.-a}

\vspace{2pc}
\noindent

\maketitle

%------------------------------------------------------------------------------------------------------------------------------------------
\bigskip
\hrule
\bigskip
%------------------------------------------------------------------------------------------------------------------------------------------
\markboth{From mindless mathematics to thinking meat?}{ }
\tableofcontents
\markboth{From mindless mathematics to thinking meat?}{ }
%------------------------------------------------------------------------------------------------------------------------------------------
\bigskip
\hrule
\bigskip
%------------------------------------------------------------------------------------------------------------------------------------------
%\clearpage
%------------------------------------------------------------------------------------------------------------------------------------------
\markboth{From mindless mathematics to thinking meat?}{}
%------------------------------------------------------------------------------------------------------------------------------------------
\def\d{{\mathrm{d}}}
\def\O{{\mathcal{O}}}
\def\omicron{o}
\def\CC{\mathbb{C}_0}
\def\N{\mathbb{N}}
\def\Z{\mathbb{Z}}
\def\Q{\mathbb{Q}}
\def\R{\mathbb{R}}
\def\C{\mathbb{C}}
%------------------------------------------------------------------------------------------------------------------------------------------
\section{Introduction}\label{S:intro}
%------------------------------------------------------------------------------------------------------------------------------------------
\parindent0pt
\parskip5pt
\vspace{-10pt}
The theme of the 2017 FQXi essay contest is this: ``Wandering towards a goal: 
How can mindless mathematical laws give rise to aims and intention?'' This is certainly an extremely interesting question, but as several other contributors have pointed out --- there are an awful lot of assumptions built into the way the question is phrased,  and in attempting to answer this question it would seem unlikely that any generally agreed-upon community-wide consensus could be reached. 
The question touches on teleology,  on emergence, (the sum is greater than its parts), on the problem of consciousness, 
and can even be argued to be connected to the collapse of the wavefunction in quantum physics. 
\vspace{-10pt}
%------------------------------------------------------------------------------------------------------------------------------------------
\section{Teleology}\label{S:teleology}
%------------------------------------------------------------------------------------------------------------------------------------------
\vspace{-10pt}
Black hole event horizons in general relativity are teleological; one has to wait till the trump of doom, and then back-track from the infinite future, to know whether or not an event horizon is present right now~\cite{observability}. Black hole event horizons can form in portions of flat spacetime, and sneak up on one unexpectedly, with \emph{zero} warning~\cite{observability}. 
There is no finite-resource experiment that any physicist could possibly perform to unambiguously detect the presence or absence of an event horizon~\cite{observability}. In contrast apparent/trapping horizons are \emph{not} teleological, and their presence or absence can, (at least for spherical horizons in spherically symmetric spacetimes), be unambiguously detected by finite-resource experiments, using quasi-local (ie, finite volume) measurements of tidal and related effects~\cite{observability}. So in a very precise Popperian sense, event horizons are simply not physics, while apparent/trapping horizons certainly are physics. 

\clearpage
So why has so much attention been paid to event horizons over the last 50 years? Simply put, event horizons make it possible to prove nice general mathematical theorems, whereas the apparent/trapping horizon variants of those theorems are much trickier and subject to qualifying technical conditions. (This is not just technical quibbling, there is real physics hiding the event horizon \emph{versus} apparent/trapping horizon distinction~\cite{thermal, ana-burning, ana-tripartite}.) Even Stephen Hawking has now abjured event horizons, not once but at least twice~\cite{Hawking-dublin, Hawking-weather}. 
Initially he did so at the Dublin meeting in 2004~\cite{Hawking-dublin}: 
\begin{quote}
	The way the information gets out seems to be that a true event horizon never forms, just an apparent horizon.
\end{quote}
More recently in 2014 Hawking asserted~\cite{Hawking-weather}:
\begin{quote}
	The absence of event horizons means that there are no black holes --- in the sense of regimes from which light can't escape to infinity.  There are, however, apparent horizons which persist for a period of time.
\end{quote}
So teleology in physics has a very mixed score-card --- the only known mainstream physics example is black hole event horizons, but while event horizons are certainly mainstream (sociological) physics, it is much less certain that they are mainstream (scientific) physics --- certainly event horizons fail Popper falsifiability, while apparent/ trapping horizons are 
perfectly acceptable science in Popper's sense. 

Again, I emphasize that these are not just technical quibbles --- the confusion regarding event horizons versus apparent/trapping horizons lies at the heart of the so-called ``information paradox'' associated with black hole evaporation, which is not in any sense a paradox when viewed in terms of Popperian-appropriate apparent/trapping horizons~\cite{ana-tripartite}.
The firewall argument simply dissolves  when viewed in terms of Popperian-appropriate apparent/trapping horizons~\cite{ana-tripartite}.

The relevance to the current essay topic is that, in the one place that teleological ideas have (sociologically) become part of mainstream physics, ultimately they have really not panned out all that well, leading to decades of unnecessary confusion.

%------------------------------------------------------------------------------------------------------------------------------------------
\section{Emergence}\label{S:emergence}
%------------------------------------------------------------------------------------------------------------------------------------------
Emergence (the sum is greater than its parts) in theoretical physics is a tricky word. There are only four real physics examples where this word makes some sort of sense:
\vspace{-10pt}
\begin{itemize}
\itemsep0pt
\item Molecular dynamics $\Longrightarrow$ fluid dynamics (truncation of the BBKY hierarchy).
\item Molecular dynamics $\Longrightarrow$ continuum solid elasticity theory.
\item Phase transitions (discontinuities that only exist in the infinite volume limit).
\item Jacobson-style thermodynamic emergent gravity. 
\end{itemize}
\vspace{-10pt}

\clearpage
Now even Jacobson-style thermodynamic emergent gravity~\cite{jacobson} is subject to some minor technical qualifications~\cite{valentina}.
(In contrast Verlinde-style entropic gravity~\cite{verlinde}  was developed some 15  years later, is non-relativistic, and cannot even successfully handle the 2-body problem~\cite{visser-on-verlinde}.)

Emergence, (in the sense of Anderson's ``more is different''~\cite{more}),  certainly applies to (and is useful in) some branches of physics, but, (and this the point for current purposes), emergence is a concept that should be used with extreme care  and discretion. When used carelessly, the word emergence degenerates into nothing more than the Stalinist aphorism ``quantity has a quality all of its own''.

 %------------------------------------------------------------------------------------------------------------------------------------------
\section{Consciousness}\label{S:consciousness}
%------------------------------------------------------------------------------------------------------------------------------------------

The problem of consciousness  (somewhat roughly) equates to the Descartes  mind-body problem: Is the mind a distinct physical entity, or is it merely an epiphenomenon, somehow emergent from the physics of the brain?  Bluntly speaking, we do not yet know enough to give any meaningful answer to this question. We may have opinions, maybe even strong opinions, on this issue --- but given what we \emph{know}, (rather than what we might like to \emph{believe}), the whole mind-body problem is still in a state of highly inconclusive emotional venting, it does not yet rise to the level of a scientific debate. 

The relevance to the current essay topic is this: The theme of the essay contest implicitly appeals to consciousness (be it human or otherwise) to even define ``aims'' or ``intentions'', and the fact that we simply do not have a coherent physical understanding of the ontology of consciousness, let alone the details of the physical mechanism by which consciousness might arise, rather undermines the whole theme of the essay contest. We  do not even have a good epistemology (a coherent framework of more or less well defined and widely agreed upon experimental protocols) by which we could address this issue.

This is not to say that valiant attempts are not being made in addressing the problem of consciousness, see for instance three books by Roger Penrose~\cite{emperor, shadows, reality},  but it is fair to say that there is no widely and generally accepted resolution of the problem of consciousness.

\enlargethispage{20pt}
%------------------------------------------------------------------------------------------------------------------------------------------
\section{Collapse of the wave-function}\label{S:collapse}
%------------------------------------------------------------------------------------------------------------------------------------------
The quite mainstream Copenhagen interpretation of quantum mechanics holds a special place for the somewhat ill-defined ``observer'', whose measurements are asserted to ``collapse the wavefunction''~\cite{bohr}.   More radical Wigner-like~\cite{friend} variants of the Copenhagen interpretation argue for the centrality of \emph{human} consciousness in the collapse process. 
The Wigner's friend argument asks this: If Wigner has a friend who undertakes the actual measurement/observation, for definiteness in a Schr\"odinger's cat experiment~\cite{anyone}, does the friend collapse the wavefunction, or is wavefunction collapse postponed until the results are communicated to Wigner?

Let me ask something significantly more pointed: What if Wigner's friend is a dog? Is dog-level consciousness, observing some quantum phenomenon, sufficient to provoke wave-function collapse. Just what level of consciousness is required for collapse?
Is cat-level consciousness sufficient? (Worse, are we really sure that  human-level consciousness is sufficient? The theologians might wish to speculate on direct angelic or demonic intervention being required to collapse the wave-function, I will steer well clear of that particular mess.)
With any degree of safety and certainty we can only conclude this: 
\begin{quote}
There is something very rotten in the eigenstate of Schr\"odinger's cat.
\end{quote}

As for other popular interpretations of quantum mechanix, it is important to realise that \emph{decoherence is not enough}.
Decoherence at best reduces quantum amplitudes to classical probabilities. But decoherence by itself  does not \emph{reify} (make real) any unique experimental outcome. Decohering Schr\"odinger's cat at best leads to 50\%--50\% classically alive/dead cat, at least no longer in a quantum superposition, but one still needs an observer to record a definite outcome. 

Many-worldism has its own issues. The original Everett variant of many-worldism treated the wave-function as a representation of our state of ignorance~\cite{everett}, (essentially as a quantum variant of Jaynes' classical maximum entropy principle~\cite{maxent1, maxent2, zipf, shannon}). More modern variants of many-worldism somewhat vaguely invoke (human?) consciousness --- with the minds of conscious entities constantly dividing down the infinitely branching yggdrasillian world-tree of future possibilities. (And, if you take the path-integral approach seriously, a merging world-tree of past possibilities.)

More subtly, one one tries to make branching universe scenarios relativistic, one seems to have to deal with non-Hausdorff manifolds; with universes splitting along the future light cones of quantum measurement events, or worse.

Again, the relevance to this essay is simply this: The major interpretations of quantum mechanix, one way or another, depend on developing an ontology and epistemology of consciousness --- and this is one reason why progress on this topic is frustratingly difficult~\cite{mystic, speakable}. (The fiendishly difficult nature of the relevant experiments is not entirely helpful either.)
Without a coherent theoretical structure to work in, the thematic ``aims and intentions'' of this essay contest are  too imprecise for useful progress.

 %------------------------------------------------------------------------------------------------------------------------------------------
\section{Thinking meat}\label{S:thinking-meat}
%------------------------------------------------------------------------------------------------------------------------------------------
We are, all of us homo sapiens, simply and literally, just thinking meat. Whether we are \emph{more} than thinking meat can (for the time being) be left to the philosophers and religious advocates; the physics community simply does not (yet) have 
appropriate techniques to have anything substantive to say on this issue.
Conversely when it comes to ``mindless mathematics'', one must consider this: Can mathematics exist without a mind to formulate it? 

This question cuts to the heart of the Platonist--Intuitionist--Constructivist debate in the philosophy of mathematics. (There are at least half-a-dozen other variants on these themes, I am simplifying~\cite{real}.)
Again we see issues of mind, and implicitly consciousness, being dragged kicking and screaming into the discussion. 
(Or maybe it is the disputants being dragged kicking and screaming into the contemplation of the need for a physical ontology and epistemology of consciousness  and the mind-body problem.)

%------------------------------------------------------------------------------------------------------------------------------------------
\section{Conclusions}\label{S:conclusions}
%------------------------------------------------------------------------------------------------------------------------------------------

In summary, the way the theme for the 2017 FQXi essay contest has been phrased, one simply has no choice but to confront issues of teleology, ontology, and epistemology --- and specifically consciousness and the mind-body problem. 
What is truly frustrating for the mathematics and physics communities, is that we do not (yet) have any suitable and appropriate well-agreed-upon mathematical/physical framework to address these issues. 
In view of this circumstance, potential resolutions of the question raised in the theme for this essay contest are unlikely (at this stage) to gain widespread support within the mathematics and physics communities. Do not give up on asking these questions, but do take cognizance of the fact that these questions are \emph{hard}; facile answers will simply not be sufficient.   
In some circles, it has recently become popular to make snide comments concerning the Popperati; but remember this --- 
whatever one's views on Popper's falsifiability criterion, it is certainly a very pragmatic and useful way of focussing attention on those ideas and techniques that are most likely to lead to interesting and widely-accepted results.

\medskip
\centerline{---\,\#\#\#\,---}

%--------------------------------------------------------------------------------------------------------------------------
\ack
%--------------------------------------------------------------------------------------------------------------------------

Supported via the Marsden Fund,  administered by the Royal Society of New Zealand.

\clearpage
%------------------------------------------------------------------------------------------------------------------------------------------
\section*{References}
%------------------------------------------------------------------------------------------------------------------------------------------

%------------------------------------------------------------------------------------------------------------------------------------------

\begin{thebibliography}{69}
%------------------------------------------------------------------------------------------------------------------------------------------

\bibitem{observability}
 M.~Visser,
  ``Physical observability of horizons'',
  Phys.\ Rev.\ D {\bf 90} (2014) 127502\\
  doi:10.1103/PhysRevD.90.127502
  [arXiv:1407.7295 [gr-qc]].
  %%CITATION = doi:10.1103/PhysRevD.90.127502;%%
  %17 citations counted in INSPIRE as of 28 Feb 2017
 
 \bibitem{thermal}
 M.~Visser,
  ``Thermality of the Hawking flux'',
  JHEP {\bf 1507} (2015) 009\\
  doi:10.1007/JHEP07(2015)009
  [arXiv:1409.7754 [gr-qc]].
  %%CITATION = doi:10.1007/JHEP07(2015)009;%%
  %17 citations counted in INSPIRE as of 28 Feb 2017
 
 
 \bibitem{ana-burning}
  A.~Alonso-Serrano and M.~Visser,
  ``On burning a lump of coal'',
  Phys.\ Lett.\ B {\bf 757} (2016) 383
  doi:10.1016/j.physletb.2016.04.023
  [arXiv:1511.01162 [gr-qc]].
  %%CITATION = doi:10.1016/j.physletb.2016.04.023;%%
  %1 citations counted in INSPIRE as of 28 Feb 2017
 
 \bibitem{ana-tripartite}
  A.~Alonso-Serrano and M.~Visser,
  ``Entropy/information flux in Hawking radiation'',\\
  arXiv:1512.01890 [gr-qc].
  %%CITATION = ARXIV:1512.01890;%%
  %3 citations counted in INSPIRE as of 28 Feb 2017
 
 
 \bibitem{Hawking-dublin}
 S.~W.~Hawking, abstract of a talk given at the GR17 conference in Dublin, Ireland, 2004.

 \bibitem{Hawking-weather}
 S.~W.~Hawking, ``Information preservation and weather forecasting for black holes'', \\
 arXiv:1401.5761 [hep-th].

 \bibitem{jacobson}
  T.~Jacobson,
  ``Thermodynamics of space-time: The Einstein equation of state'',\\
  Phys.\ Rev.\ Lett.\  {\bf 75} (1995) 1260
  doi:10.1103/PhysRevLett.75.1260
  [gr-qc/9504004].
  %%CITATION = doi:10.1103/PhysRevLett.75.1260;%%
  %1025 citations counted in INSPIRE as of 28 Feb 2017
 
 \bibitem{valentina}
 V.~Baccetti and M.~Visser,
  ``Clausius entropy for arbitrary bifurcate null surfaces'',\\
 \leftline{Class.\ Quant.\ Grav.\  {\bf 31} (2014) 035009
  doi:10.1088/0264-9381/31/3/035009
  [arXiv:1303.3185 [gr-qc]].}
  %%CITATION = doi:10.1088/0264-9381/31/3/035009;%%
  %2 citations counted in INSPIRE as of 28 Feb 2017
 
 \bibitem{verlinde}
 E.~P.~Verlinde,
  ``On the origin of gravity and the laws of Newton'',
  JHEP {\bf 1104} (2011) 029\\
  doi:10.1007/JHEP04(2011)029
  [arXiv:1001.0785 [hep-th]].
  %%CITATION = doi:10.1007/JHEP04(2011)029;%%
  %589 citations counted in INSPIRE as of 28 Feb 2017
 
 \bibitem{visser-on-verlinde}
 M.~Visser,
  ``Conservative entropic forces'',
  JHEP {\bf 1110} (2011) 140
  doi:10.1007/JHEP10(2011)140
  [arXiv:1108.5240 [hep-th]].
  %%CITATION = doi:10.1007/JHEP10(2011)140;%%
  %21 citations counted in INSPIRE as of 28 Feb 2017
  
  \bibitem{more}
  P. W. Anderson, ``More is different'', Science (new series) {\bf 177} (1972) 393--396.
 
 \bibitem{emperor}
 Roger Penrose, \emph{The Emperor's New Mind}, (Oxford, 1989).
 

  \bibitem{shadows}
   Roger Penrose, \emph{Shadows of the Mind}, (Oxford, 1994).
   
 \bibitem{reality}
  Roger Penrose, \emph{The Road to Reality}, (Alfred Knopf, 2004).
  
  \bibitem{bohr}
   N. Bohr, ``The quantum postulate and the recent development of atomic theory'', \\
   Nature {\bf 121} (1928) 580--590, 
   doi :10.1038/121580a0
   
 \bibitem{friend}
  E.P. Wigner, ``Remarks on the mind-body question'', \\
  in: I.J. Good,  \emph{The Scientist Speculates}, (Heinemann, London, 1961)
  
  \bibitem{anyone}
  ``Has anyone seen that bastard Schr\"odinger?"
 
 \bibitem{everett}
  Hugh Everett, ``Relative State Formulation of Quantum Mechanics'', \\
  Reviews of Modern Physics {\bf 29} (1957) 454--462. 
  %Bibcode:1957RvMP...29..454E. 
  doi:10.1103/RevModPhys.29.454.
 
 \bibitem{maxent1}
E.T.  Jaynes, ``Information Theory and Statistical Mechanics'',\\
 Physical Review. Series II. {\bf106} (1957) 620--630. 
 %Bibcode:1957PhRv..106..620J. 
 doi:10.1103/PhysRev.106.620. %MR 87305.
  
  \bibitem{maxent2}
E.T.  Jaynes, ``Information Theory and Statistical Mechanics II'', \\
Physical Review. Series II. {\bf108} (1957) 171--190. 
%Bibcode:1957PhRv..108..171J. 
doi:10.1103/PhysRev.108.171. %MR 96414.
 
 \bibitem{zipf}
 Matt Visser, ``Zipf's law, power laws, and maximum entropy'',
{\em New J. Phys.} {\bf 15} (2013) 043021.\\
 doi:	10.1088/1367-2630/15/4/043021
 
 \bibitem{shannon}
 Valentina Baccetti and Matt Visser, ``Infinite Shannon entropy'', \\
 	J. Stat. Mech. {\bf 2013} (2013) P04010

 \bibitem{mystic}
 \emph{Quantum questions: Mystical writings of the world's great physicists},
 edited by Ken Wilbur, (Shambhala, Boston, 1984 \& 2001)
 
 \bibitem{speakable}
 J. S. Bell, \emph{Speakable and unspeakable in quantum mechanics}, (Cambridge, England, 1987--2013)
 
\bibitem{real}
Matt Visser, ``Which number system is ``best'' for describing empirical reality?'', \\
FQXi 2012 essay contest: \\``Questioning the foundations. Which of our basic physical assumptions are wrong?''
arXiv:1212.6274 [math-ph]


%------------------------------------------------------------------------------------------------------------------------------------------
\end{thebibliography}
\end{document}